\documentstyle[12pt]{article}
\newcommand{\be}{\begin{equation}}
\newcommand{\ee}{\end{equation}}
\newcommand{\bea}{\begin{eqnarray}}
\newcommand{\eea}{\end{eqnarray}}
\begin{document}

\begin{center}

{\large\bf ODD BIHAMILTONIAN STRUCTURE OF NEW SUPERSYMMETRIC 
N =2,4 KORTEWEG DE VRIES EQUATION AND ODD SUSY VIRASORO - LIKE ALGEBRA}

\vspace{1.5cm} 

 Z. Popowicz \footnote{E-mail: ziemek@ift.uni.wroc.pl}
\vspace{1.0cm} \\
{Institute of Theoretical Physics, University of Wroclaw \\
pl.M.Borna 9, 50-204 Wroclaw, Poland}  \vspace{1.5cm}
\end{center}

\noindent{\bf Abstract.}
The general method of the supersymmetrization of the soliton equations with 
the odd (bi) hamiltonian structure is established. New version of the 
supersymmetric N=2,4 (Modified) Korteweg de Vries equation is 
given, as an example. The second odd Hamiltonian operator of the SUSY KdV 
equation generates the odd N=2,4 SUSY Virasoro - like algebra.
\vspace{2cm}

\section{Introduction}
The Kadomtsev-Petviashvili (KP) hierarchy of integrable soliton nonlinear 
evolution equations [1,2,3] is among the most important physically relevant 
integrable systems. One of the main reasons for the interest in the  
KP hierarchy in the last few years originates from its deep connection with 
the matrix models providing non - perturbative formulation of 
string theory [4]. The relations between integrable models and 
conformal symmetries have been studied intensely since 1984 [5,6]. 
Qutie recently a new class of integrable systems motivated by the 
Toda field theory appeared both in the mathematical [7] and in 
the physical literature [8].  

On the other side the applications of the supersymmetry (SUSY) to the 
soliton theory provide us a possibility of the  generalization of the 
integrable systems. The supersymmetric integrable equations [8-20] 
have drown a lot of attention in recent years for a variety of reasons. 
In order to get a supersymmetric theory we 
have to add to a system of k bosonic equations kN fermions and k(N-1) 
boson fields (k=1,2,..N=1,2,..) in such a way that the final theory 
becomes SUSY invariant. Interestingly enough, the supersymmetrizations, 
leads to new effects (not present in the bosonic soliton's theory).  
In particular the roots for the SUSY Lax operator are not uniquely defined 
[15,20], there is no bosonic reduction to the classical equations [21] and  
the non-local conservation laws [22] appeare. These effects rely 
strongly on the hamiltonian description of the supersymmetrical integrable 
systems.

In this paper we would like to describe a new method of the 
$N=2$ supersymmetrization of the Modified Korteweg 
de Vries (MKdV) and  Korteweg de Vries (KdV) equations. 
As the result our super - equations are connected with the complex version 
of the soliton equations. For this reason  this method could be compared 
with the  supercomplexification. We show that it is possible to construct, 
for the SUSY MKdV equation, the supersymmetric Lax operator.
This operator however does not generate the 
supersymmetric conserved currents defined in the whole superspace. We 
introduce  new Miura transformations which allow to supersymmetrize the KdV 

The SUSY KdV  equation, considered in this paper  constitue a
bi-hamiltonian equation and is completely integrable.
The Poisson tensors and the hamiltonians of the supercomplexified version 
of the MKdV and KdV are  superfermionic (odd) objects. The second 
Hamiltonian operator of the supercomplexified KdV equation generates 
some odd $N=2$ SUSY algebra in contrast to the even algebra (SUSY $N=2$ 
Virasoro algebra) of the supersymmetric version of the KdV equation 
considered in [12,14]. We use therefore the name {\it the odd SUSY Virasoro 
- like algebra} in order to distinguish it from the even SUSY Virasoro 
algebra of [12,14].

This supercomplexification is a general method 
and could be applied to  the $N=2$ SUSY case as well. Indeed, in this case, 
we obtained a new $N=4$ SUSY KdV equation with an odd bihamiltonian 
structure. The model proposed in this paper, differs from the
one considered in [14,17].

The idea of introducing an odd hamiltonian 
structure is not new. Leites noticed almost 20 years ago [23], that in the 
superspace one can consider both  even and odd  sympletic structures, with 
even and odd Poisson brackets respectively. The odd brackets (also known as 
antibrackets) have recently drown some interest in the context of BRST 
formalism in the Lagrangian framework [24], in the supersymmetrical quantum 
mechanics [25], and in the classical mechanics [26,27]. The present paper 
proposes some new examples of the odd sympletic structure  with  the 
$N=2,4$ supersymmetry.

The paper is organized as follows: In Section 2 we give a brief 
descriptions of the (complex) (M)KdV  equations. In Section 3 the 
basic supersymmetric notation is introduced. In Section 4 we 
present the supercomplexification of the MKdV equation. In particular 
the  Lax formulation, the odd bihamiltonian structure,  
and the superfermionic conserved currents are described. Section 5 
contains supercomplexifications of the KdV equation. We describe the 
odd bihamiltonian structure and define the hereditary recursion operator. 
The realization of the odd SUSY Virasoro - like algebra in terms of the 
solutions of the supercompexified KdV equation is given. The last section 
contains the supercomplexification of the $N=2$ SUSY KdV equation. 
 
\section{Complex (Modified) Korteweg de Vries Equation} 
The MKdV equation 
   
\be 
h_{t} = - h_{xxx} + 6h^{2}h_{x},
\ee
has the following Lax representation 
\be
L=\partial^{2} + 2 h\partial,  {~~}{~~}  
\frac{\partial}{\partial t} L =4 \Big [ L , L^{\frac{3}{2}}_{\geq 1} \Big ].
\ee
\noindent where $\geq 1$ denotes the projection onto the purely differential 
part of pseudodifferential element. 

The MKdV equation constitute the bihamiltonian system 

\be
h_{t} = P_{1} \frac{\delta H_{2}}{\delta h} = P_{2} \frac{\delta H_{1}}
{\delta h},
\ee
\noindent where 
\be
P_{1} = \partial , {~~~~}{~~~} 
P_{2} = -\partial^{3} + 4\partial h \partial^{-1} h \partial,
\ee
and 
\be
H_{1} =\frac{1}{2} \int dx h^{2},{~~}{~~}
H_{2}=\frac{1}{2} \int dx \Big( h^{4} -h_{xx}h \Big ).
\ee
The conserved currents for the MKdV equation can be obtained using the 
trace formula 
\be 
H_{n} = Tr L^{\frac{2n+1}{2}},
\ee
\noindent where $Tr$ denotes the coefficients standing in the 
front of $\partial^{-1}$ in the pseudodifferential element 
$L^{\frac{2n+1}{2}}$. 
In the following we shall also use the third conserved current
\be
H_{3} = \frac{1}{2}\int dx \Big ( h_{xxxx}h + 2h^{6} - 5h_{x}^{2}h^{2} - 
5h_{xx}h^{3} \Big ).
\ee
The complex MKdV equation can be obtained from the 
MKdV equation assuming that  $h=g + if$, where $g=g(x,t), f=f(x,t)$ 
are new functions. Inserting this expansion in the MKdV equation 
we obtain 
\be
g_{t} = \partial \Big( -g_{xx} + 2g^{3} -6f^{2}g \Big),
\ee
\be
f_{t} = \partial \Big( -f_{xx} -2f^{3} + 6g^{2}f \Big).
\ee
Our complex MKdV equations constitute the bi-hamiltonian system. This 
structure could be extracted from the usual bi-hamiltonian structure of MKdV 
equation using expansion of $h=g+if$. 
We have twice as much currents in the complex case. They
are given as the real and the complex part of (5,7). For 
the first two conserved currents (5) we have

\be
H_{1r} = \int dx (g^2 - f^2), {~~}{~~}
H_{1z} = \int dx (gf) , 
\ee

\bea
H_{2r} &=& \int dx \Big (g_{xx}g - f_{xx}f + g^{4} - 6g^{2}f^{2} + 
f^{4}\Big ), \cr 
H_{2z} &=& \int dx \Big (2g_{xx}f + 4g^{3}f -4gf^{3}\Big ).
\eea

The  MKdV equation is related to the famous KdV equation 
\be
w_{t}:=-w_{xxx} +6ww_{x},
\ee
\noindent by the Miura transformation
\be 
w:=h_{x} + h^{2}.
\ee

The bihamiltonian structure of the KdV equation is
\be
w_{t}=P_{1} \frac{\delta H_{2}}{\delta w} = P_{2} \frac{\delta 
H_{1}}{\delta w},
\ee
where 
\be 
H_{2} :=\frac{1}{2}\int dx \Big (w_{x}^2+2w^{3}\Big ) , {~~} H_{1} 
:=\frac{1}{2} \int dx w^{2},
\ee
\be
P_{1} := \partial , {~~}{~~}
P_{2} := -\partial^3 + 2\partial w + 2 w\partial.
\ee
$P_{2}$ is the so called second Hamiltonian operator which generates
the Virasoro algebra [5,6]. It is connected with 
the first Hamiltonian operator of the MKdV equation via [3]

\be
P_{2}:=\frac{\delta w}{\delta h}{~~} \partial {~~} (\frac{\delta w}{\delta 
h})^{\star},
\ee
where $\star$ denotes the hermitian conjugation, $\frac{\delta w}{
\delta h}$ is the Frechet [28] derivative, and $w$ is defined by (13).

The complex KdV equation can be obtained from the KdV equation assuming 
that $ w=s+iu $ where $s$ and $u$ are new functions. 
One gets the system of two equations
\bea
s_{t} &=& \partial \Big ( - s_{xx} + 3s^{2} - 3u^2 \Big ), \ \\
u_{t} &=& \partial \Big ( - u_{xx} +6su \Big ).
\eea
Using the same expansion of $w$ it is also possible to obtain the Lax 
representation and the bi-hamiltonian structure of the complex KdV 
equation. We have the following bi-ha\-mil\-to\-nian structure of the 
equations (18-19):
\bea
 \frac{d (s,u)^{t}}{d t} &&:=
\pmatrix { \partial & 0 \cr
	   0 & -\partial } \left(\frac{\delta H_{2r}}{\delta s},                 
     \frac{\delta H_{2r}}{\delta u}\right)^{t} = \cr
&& \pmatrix { -\partial^{3} + 2\partial s + 
2s\partial &  2u \partial + 2\partial u  \cr
 2u\partial + 2\partial u  & \partial^{3} - 2\partial s - 2 s\partial } 
	\left(\frac{\delta H_{1r}}{\delta s},
	 \frac{\delta H_{1r}}{\delta u}\right)^{t},
\eea
where $( , )^{t}$ denotes the transposition, while $H_{1r}$ and $H_{2r}$ 
are the real parts of the hamiltonians (15). 

\section{Supersymmetric Notations}

We shall consider an $N=2$ superspace with the space coordinates $x$ and 
the Grassman coordinates 
$\theta_{1},\theta_{2},\theta_{2}\theta_{1}=-\theta_{1}\theta_{2},
\theta_{1}^{2}=\theta_{2}^{2}=0$. 
The so called extended supersymmetry is assumed for which the superfields 
are superfermions or superbosons. Their Taylor expansions with respect 
to $\theta$ take the form 
\be
U(x,\theta_{1},\theta_{2}) = u_{0}(x) + \theta_{1}\xi_{1}(x) +
	\theta_{2}\xi_{2}(x) + \theta_{2}\theta_{1}u_{1}(x),
\ee
\noindent  where the fields $u_{o},u_{1}$ are interpreted as bosons 
(fermions) for the superboson (superfermion) field, while $\xi_{1},\xi_{2}$ 
as fermions (bosons) for the superboson (superfermion) field, respectively. 
The supersymmetric covariant derivatives are defined by 
\be 
\partial = \frac{\partial}{\partial x} ,{~~}{~~} D_{1} = \frac{\partial}{
\partial \theta_{1}} + \theta_{1}\partial , {~~}{~~}
D_{2} = \frac{\partial}{\partial \theta_{2}} + \theta_{2}\partial ,
\ee
with the properties
\be
D_{1}^{2}=D_{2}^{2} = \partial, {~~}{~~} D_{1}D_{2} + D_{2}D_{1} = 0.
\ee
\be
D_{1}^{-1}:= D_{1}\partial^{-1}, {~~} D_{2}^{-1} := D_{2}\partial^{-1},
\ee
\noindent where $\partial^{-1}$ is defined as the formal series
\be
\partial^{-1} f =f\partial^{-1} -f_{x}\partial^{-2}+f_{xx}\partial^{-3} - 
...,
\ee

We shall use below the following notation: $(D_{i}F)$ denotes the outcome 
of the action of the superderivative on the superfield $F$, while $D_{i}F$
denotes the action itself.

We define the integration over the $N=2$ superspace to be 
\be 
\int dX H(x,\theta_{1},\theta_{2}) = \int dx d\theta_{1}d\theta_{2}
H(x,\theta_{1},\theta_{2}),
\ee
where Berezin's convention are assumed
\be
\int d\theta_{i} \theta_{j} := \delta_{i,j}, {~~} \int d\theta_{i}:=0. 
\ee
We always assume that the components of the superfileds and 
their derivatives vanish rapidly enough. Then 
\be
\int dX (D_{1}U) = \int dX \theta_{2}\theta_{1}\xi_{1x} = \int dx \xi_{1x} 
=0,
\ee
and similarly for $(D_{2}U)$ where $U$ is an arbitrary superfunction 
vanishing at $\pm\infty$.

The $N=2$ pseudo-superdifferential operators 
\be
P:= \sum_{-\infty}^{\infty}\Big( B_{n}^{1} +F_{n}^{1}D_{1} +
	F_{n}^{2}D_{2} + B_{n}^{2}D_{1}D_{2} \Big),
\ee
\noindent where $B_{n}^{i}$ are superbosons, and $F_{n}^{i}$ are  
superfermions and $(i=1,2)$, form an associative algebra.

\section{Supercomplexifications of  N=2 MKdV equation}
The supersymmetric MKdV obtained from the Lax 
pair representation 
\be
L:=\partial^{2} + 2(D_{1}D_{2}U)\partial + 2U_{x}D_{1}D_{2}.
\ee
\be
\frac{\partial L}{\partial t}= 4\Big [ L, L^{\frac{3}{2}}_{\geq 1} \Big ],
\ee
reads 
\be 
U_{t} = -U_{xxx} -2U_{x}^{3} + 6(D_{1}D_{2}U)^{2}U_{x}.
\ee
where  $\geq 1$ denotes the projection onto 
the purely superdifferntial part of the pseudo\-super\-sym\-metric element.

This equation has several surprising properties. 
First, let us notice that introducing new superfields 

\be
g: = (D_{1}D_{2}U), {~~}{~~}
f: = U_{x},
\ee

\noindent the supersymmetric MKdV equation can be rewritten as 
the complex MKdV equation (8-9). This supersymmetric MKdV equation is 
however not equivalent with the complex MKdV since we have to take into 
account the constraint (33).

The bosonic components of the SUSY MKdV equation lead to the following 
system of  equations $(U=u_{0} + \theta_{1}\xi_{1} + \theta_{2}\xi_{2} + 
\theta_{2}\theta_{1}u_{1})$

\bea
u_{0t} &=& - u_{0xxx} - 2u_{0x}^{3} + 6u_{1}^2u_{0x} , \ \\
u_{1t} &=& \partial \Big ( -u_{1xx} + 2u_{1}^3-6u_{0x}^2u_{1} \Big ).
\eea

Notice that the boson fields do not 
interact with the fermion ones because there are no such in 
(34-35). However, the fermion fields interact with boson fields 
\bea
\xi_{1t} &=& -\xi_{1xxx} + 6\xi_{1x}u_{1} + 6\xi_{2}u_{0x}, \\
\xi_{2t} &=& -\xi_{2xxx} -6\xi_{1x}u_{0x} + 6\xi_{2x}u_{1}.
\eea

This way of the supersymmetrization we shall call in the following the  
supercomplexification. The method is based on the following 
trick: we first replace the classical function in the equations of motion 
by $(D_{1}D_{2}F) + iF_{x}$ where $F$ is some $N=2$ superfield, then  
we extract the equation on $F$ and the (bi) hamiltonian structure with its 
hamiltonians.

The most surprising property of the supercomplexified MKdV equation 
is that the conserved currents of the complex MKdV equation (10-11) are not 
conserved currents of the supersymmetric MKdV (32) equation if they are 
defined in the whole superspace. Indeed if we change the usual integration 
in (10-11) to superintegration $(dx \rightarrow dX)$ and assume the 
constraint (33) then the currents (10-11) vanishes.  

It should be stressed, that the classical currents are conserved 
quantities for the supercomplexified MKdV equation if they are defined only 
in the  usual space. In order to solve the problem of conserved currents 
defined in the whole superspace we have checked that our system does not 
have any superbosonic conserved currents up to fifth conformal dimension. 
It was verified  using the computer algebra [29,30] and assuming the 
most general form for currents. The Lax operator (30) does not produce any 
superbosonic conserved currents also. 

This observation may suggest that this system is not  (bi) hamiltonian 
on the superspace. 
There is still a possibility of a superfermionic Hamiltonian operator and 
superfermionic conserved currents. It turns out that such situation occure 
in our case. One can  construct two superfermionic currents 
\bea
H_{14} = \int dX && \Big( (D_{1}U_{xxx})U -
3(D_{1}U_{x})(D_{1}D_{2}U)^2U + 
3(D_{1}U_{x})U_{x}^{2}U \cr 
&&{~~}{~~}{~~} -6(D_{2}U_{x})(D_{1}D_{2}U)U_{x}U\Big) , \\
H_{24} = \int dX && \frac{1}{2} \Big( (D_{2}U_{xxx})U - 
3(D_{2}U_{x})(D_{1}D_{2}U)^{2}U + 3(D_{2}U_{x})U_{x}^2U +\cr
&&  {~~}{~~}{~~} 6(D_{1}U_{x})(D_{1}D_{2}U)U_{x}U\Big) ,
\eea

\noindent which generates the supercomplexified MKdV equation 
\be
U_{t} = D_{1}^{-1} \frac{\delta H_{14}}{\delta U} = 
 D_{2}^{-1} \frac{\delta H_{24}}{\delta U}. 
\ee
The Hamiltonian operator $D_{1}^{-1}$ or $D_{2}^{-1}$ defines a 
closed two-form $\Omega(D^{-1}_{i}), i=1,2$ 
\be
\Omega(D^{-1}_{i})(a,b) := \int dX a D_{i}^{-1} b = - 
\Omega(D^{-1}_{i})(b,a).
\ee

\noindent where  $a$ and $b$ are arbitrary odd vector fields. 

Although  the Hamiltonian operators $D_{1}^{-1}$ and $D_{2}^{-1}$ are 
not $O(2)$ invariant in the superspace, (as the hamiltonians 
$H_{14}$ and $H_{24}$), they are superpartners.  One can restore 
the $O(2)$ invariance  considering their linear combinations.
The operator $P_{1}:= D_{1}^{-1} - D_{2}^{-1}$ is invariant 
under $O(2)$ transformation and generates the same SUSY MKdV equation with 
the hamiltonian $H_{14}-H_{24}$.   

There are two different "second"  Hamiltonian operators for 
the supercomplexified MKdV equation:

\bea
P_{12} &:=& -D_{1}\partial - 4(D_{1}D_{2}U)\partial^{-1}U_{x}D_{2} -
4U_{x}\partial^{-1}(D_{1}D_{2}U)D_{2} \cr
&& +4(D_{1}D_{2}U)\partial^{-1}(D_{1}D_{2}U)D_{1} 
-4U_{x}\partial^{-1}U_{x}D_{1}, \ \\
P_{22} &:=& -D_{2}\partial + 4(D_{1}D_{2}U)\partial^{-1}U_{x}D_{1} +
4U_{x}\partial^{-1}(D_{1}D_{2}U)D_{1} \cr
&& + 4(D_{1}D_{2}U)\partial^{-1}(D_{1}D_{2}U)D_{2} -
4U_{x}\partial^{-1}U_{x}D_{2}.
\eea
The proof that these operators define a closed two - form is postponed to 
the next  section. 

Let us now explain the derivation of  the superfermionic conserved 
currents (38-39). We assume, in the first step of the calculations, the 
ansatz (33) and use the special computer program [29,30] \footnote{A new 
version of the program described in [30], allowing  integrating the 
superfunctions, will appeare in the new edition of Reduce 3.7 April 15 
1999.} in order to compute the usual integral of the conserved currents 
of the complex MKdV equation. The next step is to compute the 
integral $(D_{1}^{-1})$ from the nonitegrable part of the first 
integration.  The integrable part yields the first series of conserved 
currents. The computation of  $(D_{2}^{-1})$ for the nonintegrable part of 
the second integration  leads to a  purely integrable result which  gives 
the second series of conserved currents. Symbolically this procedure could 
be expressed as follows:

\be
\int dx H \Rightarrow  K_{0} + \int dx K_{1}, 
\ee
\be
(D_{1}^{-1} K_{1})  \Rightarrow  H_{1} + (D_{1}^{-1}S_{1}), {~~}{~~}
(D_{2}^{-1} S_{1})  \Rightarrow   H_{2} .
\ee

If we change the order of supersymmetric integration in (45) then 
the new $H_{1}$ and $H_{2}$ are equivalent to the old one, modulo the 
superintegrable term. 

The first classical conserved current of the complex MKdV equation yields

\be
H_{12} := \int dX (D_{1}U)U_{x}, {~~}{~~} H_{22} := \int dX (D_{2}U)U_{x}. 
\ee

Our convention for indices of  $H_{ij}$ is as follows: $j$ 
denotes the conformal dimension of the classical current, and $i = 
1,2$ denotes the index of the supersymmetric integrations.
For the second classical current we got the formulae (38-39) while 
for the third classical current we got

\bea
H_{16} &&:= (D_{1}U)\Big ( -(D_{1}D_{2}U_{xxxx}) - 
5(D_{1}D_{2}U_{xx})U_{x}^{2} + 5(D_{1}D_{2}U_{xx})(D_{1}D_{2}U)^{2} \cr
&& + 5(D_{1}D_{2}U_{x})^{2} -10(D_{1}D_{2}U_{x})U_{xx}U_{x} 
-2(D_{1}D_{2}U)^{5}  +20(D_{1}D_{2}U)^{3}U_{x}^{2} \cr 
&& -10(D_{1}D_{2}U)U_{xxx}U_{x} -5(D_{1}D_{2}U)U_{xx}^{2} - 
10(D_{1}D_{2}U)U_{x}^{4} \Big ).
\eea

The superpartner  $H_{26}$ can be obtained from  (47) using 
the $O(2)$ transformation.

\section{Supersymmetric N=2 KdV Equation, Odd Bihamiltonian 
Structure, Odd Virasoro - like algebra} 

The supersymmetric generalizations of the KdV equation based on the 
supersymmetric version of the Virasoro algebra were considered in 
[11-16]. The Hamiltonian operator generating this super equation is
\be
P_{2}:=D_{1}D_{2}\partial + 2\partial \Phi + 2\Phi \partial - D_{1}\Phi D_{1}
- D_{2}\Phi D_{2},
\ee
where $\Phi=\Phi(x,\theta_{1},\theta_{2})$ is a superboson function. The 
SUSY $N=2$ KdV equation is the following one  parameter 
family of super-Hamiltonian evolution equation 
\bea
\Phi_{t} &=&  P_{2} \frac{\delta}{\delta \Phi} \frac{1}{2} \int dX \Big ( 
\Phi(D_{1}D_{2}\Phi) + \frac{\alpha}{3} \Phi^{3} \Big ) \ \\
&=& \partial \Big ( -\Phi_{xx} +3 \Phi(D_{1}D_{2}\Phi) + \frac{1}{2} (\alpha 
-1)(D_{1}D_{2}\Phi^{2}) + \alpha \Phi^{3} \Big ),
\eea
\noindent where $\alpha$ is an arbitrary constant. It was show in 
[12,13,16] that only for three values of the parameter  $\alpha=-2,1,4$ this 
system is integrable and possesses the Lax representation. The bihamiltonian 
structure of this generalization was considered in [15]. 

We show below that it is possible to construct a new bihamiltonian 
supersymmetric generalization of the KdV equation  using our 
supercomplexified method. The result is
\be
W_{t} := - W_{xxx} + 6(D_{1}D_{2}W) W_{x},
\ee
\noindent where $W$ is the superbosonic function. 

The supersymmetric generalization of the MKdV equation considered in the 
previous section is connected with this SUSY KdV equation by the following 
Miura transformation
\be 
W:=U_{x} +2\int dx (D_{1}D_{2}U)U_{x}.
\ee

The bosonic limit of (51) in which 
$W=w_{0}+\theta_{2}\theta_{1}w_{1}$  yields the classical KdV 
equation when $w_{0}=0$. Moreover introducing 
\be
s := (D_{1}D_{2} W) ,{~~}{~~}
u := W_{x}, 
\ee
the  SUSY KdV equation (51) reduces to the complex KdV equation  
(18-19). However, due to the constraint (53) this  super 
equation and the complex KdV equation are not equivalent to each other.

One may expect that our 
SUSY KdV equation should share the same properties as the SUSY MKdV 
equation. Indeed from the knowledge of the first hamiltonian structure of 
the supercomplexified  MKdV and the Miura transformation (52) it is 
possible to obtain two different "second" Hamiltonian operators

\bea
P_{21} := \frac{\delta W}{\delta U} D_{1} \partial^{-1} (\frac{\delta W}{
\delta U})^{\star} ={~~~~~~~~~~~~} {~~~~~~~~~~~~~~~~~~~~} \cr
 D_{1}\partial  -2\partial^{-1}(D_{1}D_{2}W)D_{1} \
 - 2(D_{1}D_{2}W)\partial^{-1}D_{1} + 2\partial^{-1}W_{x}D_{2}  
+2W_{x}\partial^{-1}D_{2} ,
\eea
\bea
P_{22} :=\frac{\delta W}{\delta U} D_{2} \partial^{-1} (\frac{\delta W}{
\delta U})^{\star}={~~~~~~~~~~~~~}{~~~~~~~~~~~~~~~~~~~}\cr
 D_{2}\partial  - 2\partial^{-1}(D_{1}D_{2}W)D_{2} \
 -2(D_{1}D_{2}W)\partial^{-1}D_{2} - 2\partial^{-1}W_{x}D_{1} -
2W_{x}\partial^{-1}D_{1}.
\eea
These operators generate the SUSY KdV equation (51)
\be
W_{t} := P_{21}\frac{\delta H_{12}}{\delta W} = P_{22}\frac{\delta H_{22}}{
\delta W }, 
\ee
where 
\be
H_{12}  = - \frac{1}{2}\int dX  W(D_{1}W_{x}) , {~~}{~~} H_{22}  = - 
\frac{1}{2} \int dX W(D_{2}W_{x}). 
\ee

It is also possible to obtain the Hamiltonian operators $P_{21},P_{22}$ in 
a different manner. For this pourpose one can supercomplexify the formula 
(20) and extend  the gradient of the hamiltonian to the whole 
superspace
\be
\frac{\delta H_{1r}}{\delta s} =  D^{-1}_{2}\frac{\delta 
H_{12}}{\delta W}, {~~}{~~}
\frac{\delta H_{1r}}{\delta u}  = D^{-1}_{1}\frac{\delta H_{12}}{
\delta W}.
\ee
where 
\be
H_{1r} = \frac{1}{2}\int dx \Big ( s^{2} - u^{2} \Big ).
\ee
Now it is easy to check that these Hamiltonian operators define 
the closed two - forms. The same is true for the supercomplexified 
bi-hamiltonians operators of the MKdV equation.

It is possible to define two "first" Hamiltonian operators:
\be
P_{11} := D_{1}^{-1} ,{~~}{~~} P_{12} := D_{2}^{-1},
\ee
generating the same equation
\be
W_{t} := P_{11}\frac{\delta H_{14}}{\delta W} = P_{12}\frac{\delta H_{24}}{
\delta W},
\ee
where 
\bea
H_{14} &:=& \frac{1}{2}\int dX \Big ( -W(D_{1}W_{xxx}) + 
2(D_{1}W)((D_{1}D_{2}W)^{2} - W_{x}^{2})\Big ), \  \\
H_{24} &:=& \frac{1}{2}\int dX \Big ( -W(D_{2}W_{xxx}) + 
2(D_{2}W)((W_{x}^{2} -(D_{1}D_{2}W)^{2} \Big ).
\eea

We can construct an $O(2)$ invariant bihamiltonian structure considering 
the linear combination of $ P_{11} \pm P_{12}, P_{21} \pm P_{22} $ with $ 
H_{12} \pm H_{22}, H_{14} \pm H_{24}$. These structures  
define the same SUSY KdV (51). 

Notice that the operators $P_{21},P_{22}$ or $P_{21} \pm P_{22}$ play the 
same role as the Virasoro algebra in the usual KdV equation. There is a 
basic difference - our Hamiltonian operators generate the odd Poisson 
brackets in the odd superspace. In order to obtain the explicit realization 
of this algebra we connect the Hamiltonian operator $P_{21} - P_{22}$ with 
the Poisson bracket \be
\{ W(x,\theta_{1},\theta_{2}) , W(y,\theta^{'}_{1},\theta^{'}_{2} \} =
\Big ( P_{21} - P_{22} \Big ) (\theta_{1} - \theta^{'}_{1})
(\theta_{2} - \theta^{'}_{2})\delta(x-y),
\ee
where
\be
W(x,\theta_{1},\theta_{2}) = w_{0} + \theta_{1}\xi_{1} + \theta_{2}\xi_{2} +
\theta_{2}\theta_{1}w_{1}.
\ee
\noindent Introducing the Fourier decomposition of 
$w_{0},\xi_{1},\xi_{2},w_{1}$ 
\be 
\xi_{j} :=\sum^{\infty}_{s=-\infty} G^{j}_{s} e^{isx},  {~~}{~~} j:=1,2,
\ee
\be
w_{0}:=i\sum^{\infty}_{s=-\infty} L_{s} e^{isx}, {~~}{~~} 
w_{1}:=\sum^{\infty}_{s=-\infty} T_{s} e^{isx} - \frac{1}{4},
\ee
in (64)  we obtain
\be
\{ T_{n},T_{m} \} = \{L_{n},L_{m}\} = \{L_{n},T_{m}\} = 0,
\ee
\bea
\{T_{n},G^{i}_{m}\} &=& (n^{2}-1)\delta_{n+m,0} + 
(-1)^{i}2\frac{n-m}{m}T_{n+m}  - 2\frac{n^2-m^2}{m}L_{n+m}, \ \\
\{G^{i}_{n},L_{m}\} &=& (n-\frac{1}{n})\delta_{n+m,0} 
+2\frac{m-n}{nm}T_{n+m} +(-1)^{i}2\frac{m^2-n^2}{nm}L_{n+m},
\eea
\bea
\{G^{i}_{n},G^{i}_{m}\} &=& (-1)^{i}2\frac{m^2-n^2}{nm}\Big( G^{1}_{n+m} +
G^{2}_{n+m}\Big ),\ \\
\{G^{1}_{n},G^{2}_{m}\} &=& -2\frac{m^2-n^2}{nm}\Big(G^{1}_{n+m} -
G^{2}_{n+m}\Big).
\eea
These formulae define the closed algebra with the graded 
Jacobi identity [10,26]
\be
\sum_{cycl(a,b,c)}(-1)^{[1+a][1+c]}\{a,\{b,c\}\}=0,
\ee
where $[a]$ denotes the parity of $a$. It is the desired odd 
Virasoro - like algebra.

Unfortunately we did not  find any  Lax representation for 
the superxomplexified KdV equation. In order to obtain higher currents 
we followed the method used in the soliton theory.  In this theory one can 
apply the so called recursion operator, constructed out of the 
bihamiltonian operators $R:=P_{2}P_{1}^{-1}$, to the derivation of the 
conserved quantities. This operator genrates the currents if it is 
hereditary [3,31]. 

We were able to construct such recursion operator for the supercomplexified 
version of the KdV equation

\be
R_{1}:=P_{21}P_{11}^{-1} , {~~}{~~} R_{2}:=P_{22}P_{12}^{-1},
\ee

\noindent geting higher superfermionic currents

\bea
H_{16} &:=& \frac{1}{6}\int dX \Big ( 3W(D_{1}W_{xxxxx}) - 20 
(D_{1}W_{x})(D_{1}D_{2}W_{xx})W -45(D_{1}W_{x})W_{x}^{2}W + \cr
&& 45(D_{1}W_{x})(D_{1}D_{2}W)^{2}W - 20(D_{1}W_{xx})(D_{1}D_{2}W_{x})W - 
20(D_{2}W_{x})W_{xxx}W \cr &&+90(D_{2}W_{x})(D_{1}D_{2}W)W_{x}W - 
 20(D_{1}W_{xxx})(D_{1}D_{2}W)W \cr 
&& - 20(D_{2}W_{xx})W_{xx}W 
-20(D_{2}W_{xxx})W_{x}W \Big ).
\eea
The superpartner  $H_{26}$ can be obtained from $H_{16}$  using 
the $O(2)$ transformation.

In general  $R_{1}$ and $R_{2}$ are hereditary operators. Clearly $P_{11}$ 
and $P_{21}$ or $P_{12}$ and $P_{22}$ are compatible. Indeed the deformation 
\be
W \Rightarrow W + \Big ( \epsilon + \theta_{1} \eta_{1} + \theta_{2} 
\eta_{2} +\theta_{2}\theta_{1} \varepsilon \Big ),
\ee
where $\epsilon, \varepsilon $ are arbitrary constants while $ 
\eta_{1},\eta_{2}$ are arbitrary grassmanian constants, maps $P_{21}$ 
or $P_{22}$ into the Hamiltonian operator $P_{21} - 4\varepsilon P_{11}$ 
or $P_{22} -4\varepsilon P_{12} $. Hence the resulting recursion operators 
$R_{1}$ and $R_{2}$ are hereditary [3,31]. In this sense the SUSY KdV 
equation (51) is completely integrable. 

\section{Supercomplexified N=4 KdV equation}
Let us now apply our supercomplexification method to the $N=2$ SUSY KdV 
equation. We consider the case $\alpha=4$ only. Other cases can be 
considered in an analogous way. We assume that the superfield $\Phi$ 
satisfying the $N=2$ SUSY KdV equation (48) takes after the 
supercomplexification the following form
\be
\Phi :=(D_{3}D_{4}\Upsilon) + i\Upsilon_{x},
\ee
\noindent where $D_{3}=\frac{\partial}{\partial \theta_{3}} + 
\theta_{3}\partial, D_{4}=\frac{\partial}{\partial \theta_{4}} + 
\theta_{4}\partial$ and $\Upsilon$ is some $N=4$ superboson field. 
Substituting this form in (48) we obtain
\bea
\Upsilon_{t} &&:= -\Upsilon_{xxx} + 3(D_{1}D_{2}\left( 
\Upsilon_{x}(D_{3}D_{4}\Upsilon)) \right) + 3\Big ( 
\Upsilon_{x}(D^{4}\Upsilon) + (D_{3}D_{4}\Upsilon)(D_{1}D_{2}\Upsilon_{x}) 
\Big) \cr
&& {~~~}{~~~}{~~}{~~}{~~} -4\Upsilon^{3}_{x} + 
12(D_{3}D_{4}\Upsilon)^{2}\Upsilon_{x}.
\eea

\noindent where $D^{4}=D_{1}D_{2}D_{3}D_{4}$. It is the desired 
generalization of the $N=4$ SUSY KdV equation, which is different from the
one considered in [14,17].

The supercomplexification of the bihamiltonians structures yields
\be
\Upsilon_{t} = P_{14} \frac{\delta H_{24}}{\delta \Upsilon}=P_{24} 
\frac{\delta H_{14}}{\delta \Upsilon},
\ee
\noindent where 
\bea
P_{14} && := D_{1}D_{2}D_{4}\partial^{-1} + 2\Big (D_{3}D_{4}\Upsilon 
\Big ) \partial^{-1}D_{4} + 2\partial^{-1}\Big (D_{3}D_{4}\Upsilon \Big ) 
D_{4} \cr
&&-D_{1}\partial^{-1}\Big (D_{3}D_{4}\Upsilon\Big ) 
\partial^{-1}D_{1}D_{4} -D_{2}\partial^{-1}\Big (D_{3}D_{4}\Upsilon \Big 
)\partial^{-1} D_{2}D_{4} \cr 
&& +2\Upsilon_{x}D_{3}\partial^{-1} + 2\partial^{-1}\Upsilon_{x}D_{3} 
-D_{1}\partial^{-1}\Upsilon_{x}\partial^{-1} 
D_{1}D_{3}-D_{2}\partial^{-1}\Upsilon_{x}\partial^{-1} D_{2}D_{3}, \ \\
H_{24} && :=\frac{1}{2}(D_{1}D_{2}D_{4}\Upsilon_{x})\Upsilon +
\frac{2}{3}(D_{3}\Upsilon)
\Big (\Upsilon^{2}_{x} - (D_{3}D_{4}\Upsilon )^{2}\Big ), \ \\
P_{24} && :=\partial^{-1}D_{4}\ \\
H_{14} && :=\frac{1}{2} \Big ( (D_{4}\Upsilon_{xxx})\Upsilon 
-6(D_{1}D_{2}D_{4}\Upsilon)\Upsilon_{x}(D_{3}D_{4}\Upsilon) \cr
&& -6(D_{4}\Upsilon)\Upsilon_{x}(D^{4}\Upsilon) - 
6(D_{4}\Upsilon)(D_{3}D_{4}\Upsilon)(D_{1}D_{2}\Upsilon_{x}) \cr
&& -6(D_{4}\Upsilon_{x})\Upsilon^{2}_{x}\Upsilon + 
6(D_{4}\Upsilon_{x})(D_{3}D_{4}\Upsilon)^{2}\Upsilon 
-12(D_{3}\Upsilon_{x})(D_{3}D_{4}\Upsilon)\Upsilon_{x}\Upsilon \Big ),
\eea

This is our bihamiltonian formulation of the equation (78). 
Using the same methods as in the previous 
sections it is possible to obtain an explicit realization of the odd SUSY 
$N=4$ Virasoro - like algebra. 

Let us remark that the supercomplexification of the $N=1$ SUSY 
KdV equation gives the even bihamiltonian structure of the $N=3$ SUSY KdV 
equation.
\vspace{0.5cm}

{\bf Acknowledgement} The author wish to thank  prof.J.Lukierski and 
dr A.Pashnev and dr Krivonos for the fruitfull discussion in the 
subject. This paper was supported by the KBN grant 2 P03B 136 16.

\end{document}